# IDeF-X ASIC for Cd(Zn)Te spectro-imaging systems.

O. Limousin, O. Gevin, F. Lugiez, R. Chipaux, E. Delagnes, B. Dirks, B. Horeau

*Abstract--* Joint progresses in Cd(Zn)Te detectors, microelectronics and interconnection technologies open the way for a new generation of instruments for physics and astrophysics applications in the energy range from 1 to 1000 keV. Even working between –20 and 20°C, these instruments will offer high spatial resolution (pixel size ranging from 300 × 300 µm² to few mm²), high spectral response and high detection efficiency. To reach these goals, reliable, highly integrated, low noise and low power consumption electronics is mandatory. Our group is currently developing a new ASIC detector front-end named IDeF-X, for modular spectro-imaging system based on the use of Cd(Zn)Te detectors. We present here the first version of IDeF-X which consists in a set of ten low noise charge sensitive preamplifiers (CSA). It has been processed with the standard AMS 0.35 µm CMOS technology. The CSA's are designed to be DC coupled to detectors having a low dark current at room temperature. The various preamps implemented are optimized for detector capacitances ranging from 0.5 up to 30 pF.

## I. INTRODUCTION

SINCE our previous development IBIS/ISGRI gamma-ray camera [1] on board the INTEGRAL Satellite, we have demonstrated the possibility to make use of a huge quantity of CdTe detectors associated with microelectronics front-end in space in safe conditions. On the other hand, progresses in CdTe detectors manufacturing (crystal quality and volume size), microelectronics and interconnection technologies open the way for a new generation of 1 to 1000 keV photon-energy-detectors for physics and astrophysics, offering high spatial resolution (pixel size: ~ 300 × 300 µm² to few mm²), high spectral response and high detection efficiency, and operating between –20 and 20°C. To reach these goals, reliable, highly integrated, low noise and low power consumption electronics is mandatory.

Our group is currently developing a new modular spectro-imaging system based on CdTe detectors coupled to a dedicated readout ASIC, named IDeF-X (Imaging Detector Front-end). This device will be assembled in large area cameras (100 to 1000 cm²) for space borne astrophysics, either on focusing telescope focal plane (e.g. SIMBOL-X [2] and MAX [3]), operating with hard X-rays (4 to 150 keV) or gamma-rays (511 and 847 keV), or on a large area position sensitive detector of a coded aperture telescope (4 to 600 keV) (e.g. ECLAIRs [4]).

In this paper, we will focus on the IDeF-X ASIC characteristics and performances. We will pursue with the response measured with one of the CSA, well suited for moderate detector capacitance (< 5 pF), connected to a set of CdTe detectors. Finally, we will present results of total ionizing dose tests performed on the chip with a $^{60}$Co source up to 224 krad.

## II. IDeF-X ASIC DESIGN

The developments of the IDeF-X front-end ASIC includes several successive chips, starting from the stand-alone preamplifier study and finishing with a 32 to 256 channels circuit for high-pixel density CdTe readout. We present here the very first version of the front-end electronics of our device.

TABLE I
IDeF-X CSA CHARACTERISTICS

| CSA # | Input transistor type | Input transistor size W/L (µm/µm) | Input capacitance range (pF) | Detector application |
|---|---|---|---|---|
| 0 | PMOS | 310/0.5 | 0.5 | NA (no pad) |
| 1 | PMOS | 1000/0.5 | 2 to 5 | Cd(Zn)Te |
| 2 | PMOS | 1550/0.35 | 5 to10 | Cd(Zn)Te |
| 3 | PMOS | 1600/0.5 | 5 to10 | Cd(Zn)Te |
| 4 | PMOS | 1400/0.75 | 5 to 10 | Cd(Zn)Te |
| 5 | NMOS | 1550/0.35 | 5 to 10 | Cd(Zn)Te |
| 6 | NMOS | 1600/0.5 | 5 to 10 | Cd(Zn)Te |
| 7 | NMOS | 1400/0.75 | 5 to 10 | Cd(Zn)Te |
| 8 | PMOS | 4000/0.5 | 30 | cooled Ge |
| 9 | PMOS | 2700/0.75 | 30 | cooled Ge |

The goal of this first chip is to evaluate the AMS 0.35 µm CMOS technology for low noise and low power consumption analogue design. Therefore, we have implemented a set of ten low noise charge sensitive preamplifiers (CSA), well suited to high energy applications. The CSA are designed to be DC coupled to detectors with a low dark current at room temperature (< 5 nA). The implemented CSA are optimized for detector capacitances from 0.5 to 30 pF (see table I).

The CSA electrical design is based on a "folded cascode amplifier" [5, 6] with either a PMOS or a NMOS input

B. Dirks, B. Horeau and O. Limousin are with the CEA Saclay DSM/DAPNIA/Service d'Astrophysique, bât. 709 L'Orme des Merisiers, 91191 Gif-sur-Yvette, France (e-mail: olimousin@cea.fr).

R. Chipaux, E. Delagnes, O. Gevin and F. Lugiez are with the CEA Saclay DSM/DAPNIA/Service d'Électronique, de Détecteurs et d'Informatique, bât. 141, 91191 Gif-sur-Yvette, France.

transistor. The value of the feedback capacitance is 300 fF and 500 fF respectively for CSA#0 to #7 and for CSA#8 and #9. The DC feedback is done by a PMOS transistor biased by the detector leakage current. Each CSA output is connected to a 10× voltage gain stage.

These stages are multiplexed toward a low output impedance buffer. All channels are connected to an input pad except CSA#0. Inputs can be connected to a test input $Ve\_test$ with an individual 300 fF and 500 fF on chip injection capacitor, respectively for CSA#0 to #7 and CSA#8 and #9.

On the other hand, in order to compensate a reverse detector current or to simulate a detector current without detector, each channel includes a tunable current source $il$ driven by the gate voltage $V_{il}$ (figure 1).

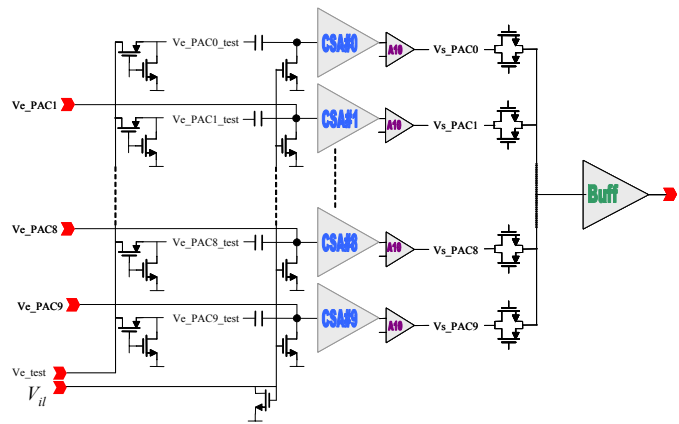

Figure 1: IDeF-X synoptic: Ten CSA's are disposed before a gain amplifier, a multiplexer directed to a low impedance output buffer. At the input, each CSA has its own test capacitor. All CSA are connected to a pad except CSA#0.

The IDeF-X layout is represented in the figure 2.

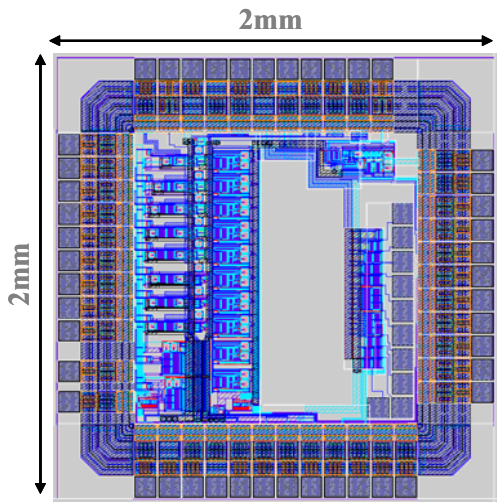

Figure 2: IDeF-X Layout. Ten preamps are disposed on the left hand side of the 4 mm² circuit.

## III. RESULTS

### A. Equivalent noise charge measurements

The first characterization of the circuit consists in measuring the Equivalent Noise Charge (ENC) of each channel as a function of the peaking time when the CSA is placed at the input of a tunable CR-RC² filter or CR-RC filter.

To perform the measurements, the circuit is packaged into a standard JLCC chip carrier and mounted on a standard printed circuit board (PCB) into the setup described below.

#### 1) ENC measurements test bench

The JLCC carrier is mounted on a test board allowing biasing, configuration, injection and response measurements. This board is inserted in an automatic ENC vs. peaking time test bench shown on figure 3 [7]. This setup includes a CR-RC² filter with tunable peaking times ranging from 20 ns up to 10 μs. It also includes a wave form generator. The pulse shapes and the noise are alternately recorded on a digital oscilloscope for the various filtering time-constants and analyzed with a computer.

ENC measurements for larger peaking times, up to 1 ms, were done manually with CR-RC filter. This last operation was not systematically done in the following.

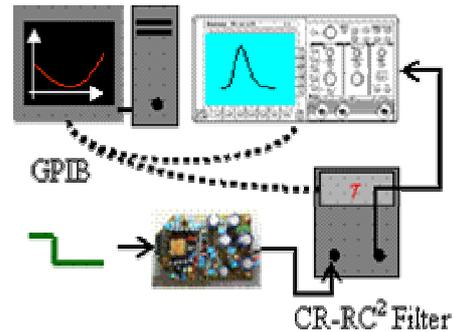

Figure 3: Setup for ENC measurement is operated with an external programmable CR-RC² shaper.

#### 2) Main ENC results for all of the CSA's

A significant part of the noise is due to the setup itself (JLCC parasitic capacitance, connectors, PCB parasitic capacitance and dielectric losses). Therefore, we have proceeded in two kinds of ENC measurements:

- In a first step, in order to measure the intrinsic performances of the circuit, the inputs were totally disconnected from the setup (no bonding on the inputs). This allows estimating the floor noise of each CSA.
- In a second step, we measured the performances with the inputs connected with a wire-bonding to the JLCC carrier. This allows using the chip with a detector. Nevertheless, we kept the JLCC input pads out of the PCB and connector. Detectors were installed directly on the JLCC by wire soldering.

We have measured the ENC for each of the ten CSA's. In the table II, we show the measured ENC values extracted for each CSA without bonding. These measurements illustrate the intrinsic performances without assumption on the setup quality. Depending on the input transistor type and size, the floor noise has been measured between 31.5 and 49.3 electrons rms.

In the table II, the CSA's were polarized with 1 mA / 3.3 V to reach the very best performances.

TABLE II
IDeF-X CSA MAIN PERFORMANCES

| CSA # | Min ENC (electrons rms) | Peaking time at ENC min (µs) |
|---|---|---|
| 0 | 12.4 (no pad) | 8.9 |
| 1 | 31.5 | 9.1 |
| 2 | 33.1 | 9.1 |
| 3 | 32.3 | 9.1 |
| 4 | 34.1 | 9.1 |
| 5 | 44.8 | 4.5 |
| 6 | 49.3 | 4.5 |
| 7 | 47.6 | 4.5 |
| 8 | 30.4 | 4.5 |
| 9 | 29.2 | 4.5 |

The noise of the CSA's has been measured with additional parasitic capacitances at the input. The measurement was operated with a 9 µs or 4.5 µs peaking time, respectively for PMOS or NMOS type CSA's, corresponding to their best noise level when no detector is connected. It led to:
- 3 to 5 electrons/pF for PMOS type CSA
- 5 to 6 electrons/pF for NMOS type CSA

In the following sections, we will concentrate on the PMOS type CSA#0 and CSA#3, and on the NMOS type CSA#6, to analyze the performances in deeper details. Since it has no anti-ESD input pad, CSA#0 is a reference for the design noise behavior without the influence of the external noise sources. The PMOS CSA#3 (best noise in the 5 to 10 pF range PMOS CSA's) and NMOS CSA#6 (same input transistor dimensions as CSA#3) are the preamps best matched to the CdTe pixel detectors typical capacitances.

*3) ENC vs. peaking time for CSA#0, #3 and #6*

When no detector is connected but when the ASIC is bonded to the chip carrier, the minimum noise level in the ENC vs. peaking time characteristics is 69 electrons rms at 9.1 µs for the PMOS type input transistor (CSA#3) and 76 electrons rms at 4.5 µs peaking time for the NMOS input transistor (CSA#6). In the latest, injection of a current *il* is necessary to make the CSA work properly. This current compensates the reverse current from the anti-ESD input diodes and adds a parallel noise contribution to the ENC vs. peaking time characteristics. This is the reason why the minimum value of the noise occurs for shorter peaking times than in the PMOS case where no additional current is mandatory.

On CSA#0, without input pad, we can measure intrinsic performances and limits of the design. Its minimum noise level in the ENC vs. peaking time characteristics is 12.4 electrons rms for a 9.1 µs peaking time.

ENC vs. peaking time characteristics are plotted for CSA#0, CSA#3 and CSA#6 in figure 4.

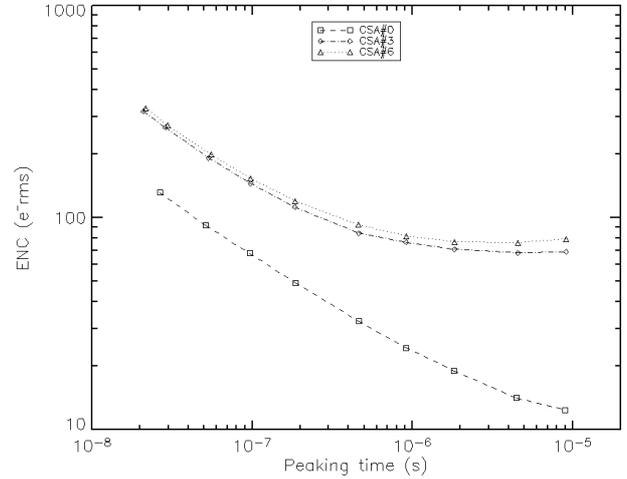

Figure 4: ENC vs. peaking time measurement results for CSA#0, CSA#3 and CSA#6. The CSA#0 has no input pad. On CSA#3 and CSA#6, the inputs are connected to the chip carrier but no detector is present. CSA's are polarized with 1mA current through the input transistor.

Forthcoming ENC evaluation on new IDeF-X versions will be achieved with the chip mounted on a ceramic board or low dielectric loss factor materials board to reach the best performances.

*4) ENC behavior with different biasing conditions*

Up to now, the results were obtained with high power consumption operations – 1 mA i.e 3.3 mW. The power consumption can be reduced by limiting the current in the input transistor but this will impact the noise. We have recorded the noise behavior of the CSA#3 as a function of the current in the input transistor (200 µA i.e. 660 µW and 50 µW i.e. 165 µW). Results are shown in figure 5.

As expected, as the total current in the CSA reduces, the serial noise increases: it is oppositely proportional to the square root of the input transistor transconductance, i.e. it is roughly oppositely proportional to the square root of the current passing through the input transistor, operating in weak inversion.

For long peaking times (above 8 µs), the parallel and *1/f* noise contributions begin to dominate and no more influence of the bias condition is visible on the ENC.

We conclude that our design can be easily used in a low consuming input stage if large peaking-times (> 8–10 µs) are applicable. This means that we need to operate detectors with leakage currents as low as possible, at least lower or equal to the internal leakage currents of the chips, mainly due to the pads. The dark current has to be less than few pico amps, as shown in section IV. A moderately cooled – around 0°C – Schottky CdTe detector or pixel CdZnTe detector can reach such values.

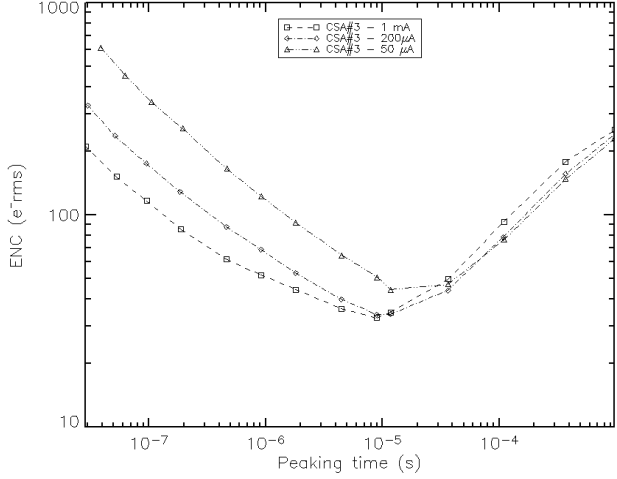

Figure 5: ENC vs. peaking time as a function of the bias current in the CSA#3 input stage of IDeF-X. The ENC are obtained on a chip without bonding at the entrance to avoid noise due to the setup conditions (PCB, JLCC, connectors contributions).

### B. Spectroscopy measurements

These promising results allowed us to consider a direct application for spectroscopy, thus we have connected the PMOS CSA#3 of IDeF-X, biased with 200 µA (660 µW), to a set of CdTe detectors at room temperature (21-24°C). The detectors were DC coupled to the input of the CSA which output was connected to a Canberra 2025 amplifier with an adjustable shaping time from 0.5 to 12 µs corresponding to 1.5 µs to 36 µs peaking time with Gaussian shaping.

First of all we have plugged a $2 \times 2 \times 2$ mm$^3$ Travelling Heater Method (THM) grown CdTe (Eurorad) equipped with quasi ohmic platinum electrodes. This detector showed a dark current of 5 nA when biased at 100 V. Its capacitance of ~0.2 pF is dominated by other parasitic and interconnections capacitances. Since the current is not negligible and the detector capacitance is very low, the lowest noise was obtained for a 0.5 µs shaping time. We had nice line shapes with a 3.5 keV FWHM at 59.5 keV and 2.2 keV FWHM at 17.8 keV (fig. 6).

We have also connected to the same CSA a $4.1 \times 4.1 \times 0.5$ mm$^3$ THM grown CdTe (ACRORAD) equipped with a Schottky contact at the anode and a guard ring at the cathode (1 mm guard ring surrounding the $2 \times 2$ mm² pixel). The reverse dark current of the diode is very low (< 10 pA under 200 V bias voltage at 21°C) and the capacitance of the pixel is 0.7 pF. The spectrum shown on figure 7 illustrates the results with 2 µs shaping time: 1.6 keV FWHM at 59.5 keV and 1.4 keV FWHM at 13.8 keV.

In the ISGRI CdTe gamma camera equipped with mixed analogue and digital ASIC [1, 8, 9], 2.8 mW power consuming, the best spectral resolution measured during the ground calibration phase, with $4 \times 4 \times 2$ mm$^3$ THM CdTe crystals (ACRORAD), biased under 100 V at 0°C, was 5.6 keV FWHM at 60 keV.

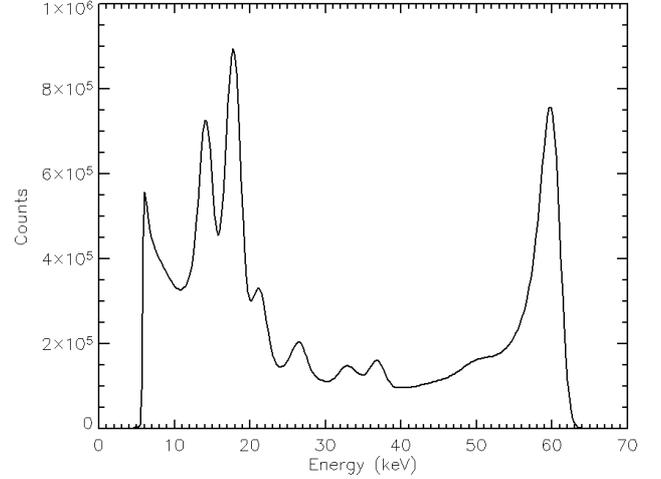

Figure 6: Spectrum of a $^{241}$Am source obtained with a $2 \times 2 \times 2$ mm$^3$ CdTe detector equipped with a Pt contacts on both electrodes (EURORAD, France) plugged on CSA#3, biased with 200 µA. The detector is biased under 100 V at 24°C. The spectral response is good (3.5 keV FWHM at 59.5 keV). The broadening on the left hand side of the 59.5 keV line is mainly due to the charge loss and ballistic deficit in the 2mm thick CdTe.

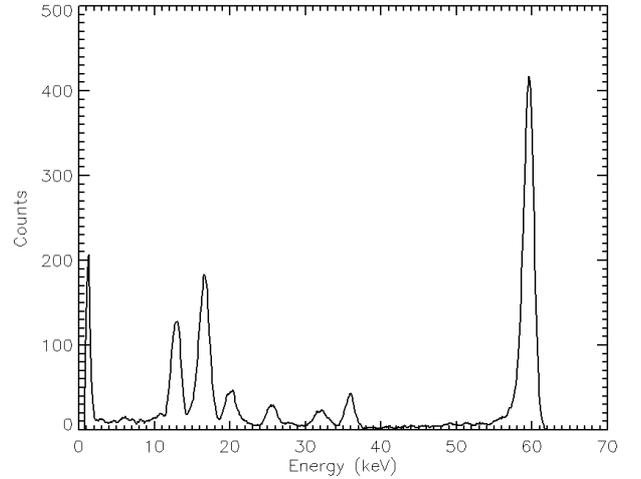

Figure 7: Spectrum of a $^{241}$Am source obtained with a $4.1 \times 4.1 \times 0.5$ mm$^3$ CdTe detector equipped with a Schottky contact at the anode (ACRORAD, Japan) plugged on CSA#3, biased with 200 µA. The cathode is $2 \times 2$ mm² pixel surrounded by a 1 mm guard ring. The detector is biased under 200 V at 21°C. The spectral response is very good (1.6 keV FWHM at 59.5 keV and 1.4 keV FWHM at 13.8 keV). The very low threshold value around 2 keV is noticeable.

## IV. IRRADIATION WITH $^{60}$CO

The successive versions of IDeF-X ASIC are devoted to future space-borne application in astrophysics and a first evaluation of the design and technology appeared necessary from the radiation point of view. Thus, we have irradiated two IDeF-X circuits with a 589 GBq $^{60}$Co source during 224 hours.

The ASIC were disposed at 13 cm from the source, leading to a 1 krad/h dose rate. The irradiation was segmented in nine steps leading to measurements at 10, 20, 30, 63, 95, 111, 151, 161 and 224 krad of cumulated dose.

One chip was used to the PMOS type CSA#3 analysis and the second was used to the NMOS type CSA#6 analysis. Both CSA's were correctly biased to ensure their functionality during the runs with a 200 µA current in the input transistor. All others CSA's in the chip were also biased but not multiplexed to the output and therefore, not systematically monitored. In the NMOS type CSA#6, a high enough current $il$ was injected in the reset transistor to compensate a potential shift of the leakage current. This helps to keep the preamp operating even with current shifts.

The spectral response of the two circuits was monitored during the nine steps of irradiation. A calibrated injected signal was sent to the ASIC inputs while we recorded the response after a 3 µs shaping time amplifier (ORTEC DUAL SPEC 855) in a standard spectroscopy chain.

Between irradiation runs, the two CSA's #3 and #6 were fully characterized (ENC, gain, rise-time and fall-time, polarization currents) with the test bench described above. For the NMOS CSA#6, the compensation current $il$ was adjusted to its optimal minimum value for the fine characterization.

### A. Results

#### 1) Amplification gain

Along the nine irradiation steps, we have followed the amplification gain of the CSA chains measuring the output voltage level directly behind the output buffer. No shift was found, neither the PMOS nor the NMOS type CSA. The amplitude was permanently measured at 37 mV and 35.5 mV respectively for the CSA#3 and #6 for a 4 mV square signal injected through the 300 fF internal injection capacitor.

#### 2) Output signal rise-time and fall-time

Likewise, the signal output rise-times have not changed at all for both kinds of CSA's. We measured 35 ns and 38 ns rise-times respectively for the CSA#3 and #6. On the other hand, we have measured slight changes of the output signal fall-times in both designs:

· In the case of CSA#3, since the current through the reset transistor is very low (~ 380 fA before irradiation), the output signal fall-time is long (> 35 ms). After 95 krad, the fall-time has decreased to 13 ms and falls down to 10 ms after 224 krad. This effect is typical of an increase of the current $I_R$ through the reset transistor. It acts as a feedback resistor $R_R$, oppositely proportional to $I_R$. After 224 krad, the current through the reset transistor is evaluated to ~1.5 pA.

· In the case of the CSA#6, we had to tune the compensation current $il$ after each run in order to reach fall-time values as long as possible. Therefore, the fall-time is not immediately representative of the CSA evolution with irradiation. We will detail this point later in the paper.

#### 3) Power consumption

At the end of the campaign, we have measured the power consumption (current level in the 3.3 V power supply) and we have concluded that absolutely no shift occurred during the irradiation. The measured current is the total current in all the circuit (Ten CSA's, amplifiers, multiplexer and output buffer). It is not possible to distinguish between the NMOS and PMOS CSA's in this case.

#### 4) ENC measurements

Next, we have systematically measured the ENC vs. peaking time characteristics of the CSA#3 and CSA#6. Figure 8 shows the evolution of the minimum ENC value as a function of the cumulated dose. This minimum value is linearly degrading with the dose and the designs seem to slowly degrade at almost the same speed (0.09 and 0.1 electrons rms/krad respectively for PMOS and NMOS type CSA). The minimum ENC value always occurs between 1.8 and 4 µs peaking time for NMOS CSA. This peaking time is constrained by the compensation current which imposes a strong parallel noise structure. But for the PMOS CSA, the minimum ENC values occurs at 9 µs peaking time until 111 krad. For a higher dose, the minimum ENC occurs at 4.5 µs peaking time which is a sign of a progressive increase of the parallel noise contribution.

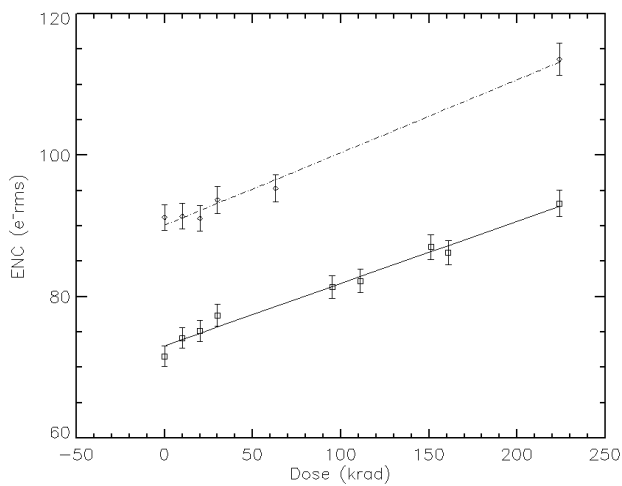

Figure 8: Minimum of the ENC vs. peaking time characteristics as a function of the absorbed dose in PMOS CSA#3 and NMOS CSA#6. Both CSA's are polarized under 200 µA / 3.3 V conditions. A linear function fits both sets of data.

Finally, after the end of irradiation, the two circuits have been annealed during two months at room temperature. After this period, no improvement or post irradiation effect has been observed and the circuit characteristics stayed remarkably stable.

### B. Discussion

In this section, we concentrate on the analysis of the noise structure for the PMOS and NMOS CSA's, before irradiation and after the 224 krad irradiation dose. First we describe the PMOS evolution because it had the lowest noise all time long.

*1) Noise structure in PMOS input CSA*

Figure 9 illustrates the ENC vs. peaking time characteristics for the PMOS CSA#3, before and after irradiation. We easily distinguish the high frequency serial noise and the low frequency parallel noise, superimposed to the *1/f* noise. Thanks to the model fitting and according to the AMS CMOS technology parameters, we can extract the total capacitance at the CSA input, the transistor transconductance, the total parallel noise and the *1/f* noise level.

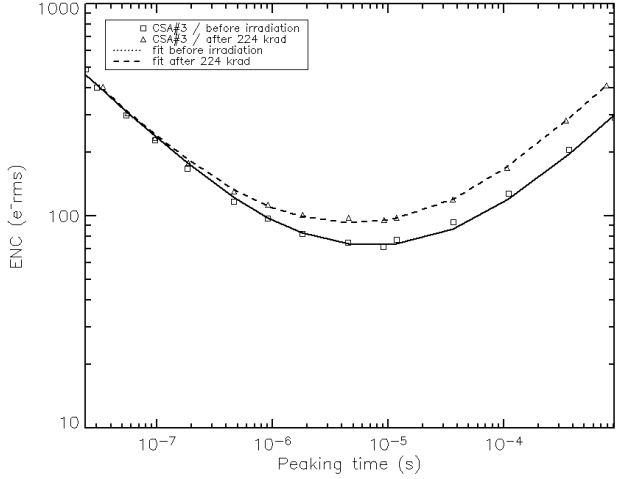

Figure 9: ENC vs. peaking time for the CSA#3 before and after 224 krad of irradiation. CSA#3 is polarized with 200µA current through the input transistor. Lines represent the noise model that fits the data.

First of all, we see that the serial noise is not affected by the irradiation which means that the transconductance of the input transistor (3.2 mS for 200 µA / 3.3V) stay unchanged with the dose. On the other hand, the total capacitance is estimated to be 6.1 pF.

At very low frequencies, the noise is totally dominated by the parallel noise contribution. It does increase with the dose. This was already visible previously by the fact that the output signal fall-time was diminishing with the dose.

The total current $I_{tot}$, responsible for the parallel noise includes the following sources represented on figure 10:
- The pad leakage current $I_1+I_2$ (two reverse bias diodes),
- The compensation current *il* driven by $V_{il}$
- The reset transistor noise, depending on $I_R = I_2 - I_1 + il$

The total current $I_{tot}$ is defined by the relation (eq. 1):

$$2q \cdot I_{tot} = 2q \cdot (I_1 + I_2 + il + I_R) \quad \text{(eq. 1)}$$

where $q$ is the electron charge.

The current $I_R$ is extracted from the CSA output signal fall-time. The current *il* is extracted from the CSA#0 output signal fall-time. As a matter of fact, CSA#0 is deprived of pad, so $I_R$ is equal to *il*.

When $V_{il}$ = 0V in the PMOS case, we found *il* ~ 200 fA before irradiation and *il* ~ 1 pA after 224 krad probably due to the threshold voltage shift of the NMOS current mirror transistors.

On the other hand, as mentioned before, the current through the reset transistor $I_R$ goes from 380 fA to 1.5 pA.

Next, the model fitting of the parallel noise allows extracting the total current $I_{tot}$. It is 8.7 pA before irradiation and rises to 17 pA after 224 krad irradiation dose.

At this level, we deduce from this analysis that the reverse current of the anti-ESD pad diodes is by far the main parallel noise contribution even if the current through the reset transistor increases more rapidly with the dose.

Finally, the ENC vs. peaking time characteristics shows a 25% augmentation of the *1/f* noise.

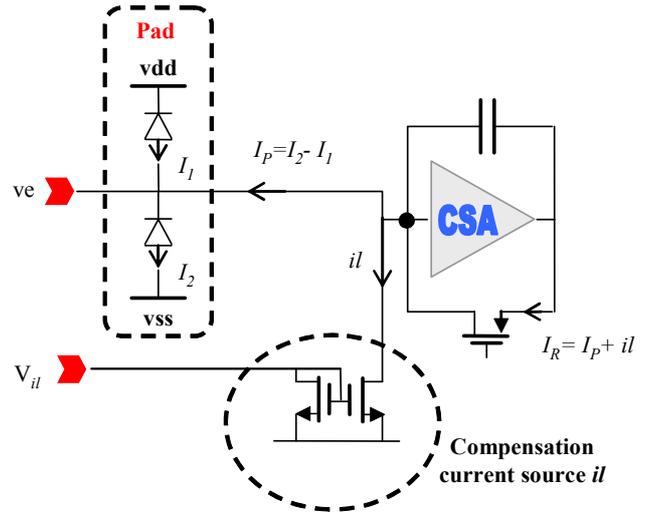

Figure 10: Scheme of the CSA. Definition of the currents: $I_1$ and $I_2$ are the reverse bias diode currents of the pad, *il* is the compensation current (or residual current when $V_{il}$ = 0 V), $I_R$ is the current through the reset transistor.

*2) Noise structure in NMOS input CSA*

Figure 11 illustrates the ENC vs. peaking time characteristics for the NMOS CSA#6, before and after irradiation.

We see that the serial noise is not affected by the 224 krad irradiation dose. Therefore, the transconductance appears stable at ~ 3 mS for 200 µA / 3.3V polarization. Note that the CSA#6 NMOS input transistor has exactly the same dimensions as the CSA#3 PMOS input transistor. Consequently, in the weak inversion region, it was predictable to see a similar value of the transconductance for the NMOS CSA#6 and the PMOS CSA#3.

At very low frequencies, the noise is dominated by the parallel noise. It increases with the dose only because it is necessary to increase the compensation current *il* to make the CSA#6 work properly.

The fact we need to impose a positive *il* current in the NMOS CSA seems to demonstrate that the current $I_P$ is negative. In the PMOS CSA, since no adjustment of *il* is necessary, $I_P$ must be positive (see figure 10 for current sign definition). The current $I_P$ is negative in the NMOS CSA probably because the gate of the input transistor has a much lower voltage (typically 0.5 V) than in the PMOS CSA (typically 2.8 V). In order to work properly, the CSA needs the current $I_R$ to remain positive. In the NMOS CSA, this implies that the compensation current $|il|$ has to stay larger than the current from the pad $|-I_P|$.

Before irradiation, the gate of the compensation current mirror (see figure 10) is set to $V_{il} = 198$ mV. After irradiation, this value goes to $V_{il} = 235$ mV.

In order to determine the corresponding values of the compensation current $il$, we have measured the output signal fall-time of the CSA#0 (no pad, i.e. the current $I_R$ in the reset transistor is equal to the compensation current $il$). CSA#0 and others have exactly the same compensation current sources, so measurements on CSA#0 are representative for others CSA on the chip. As mentioned before, the fall-time is directly linked to the current through the reset transistor.

We found that $il$ was ~ 40 pA before irradiation and ~ 80 pA after irradiation.

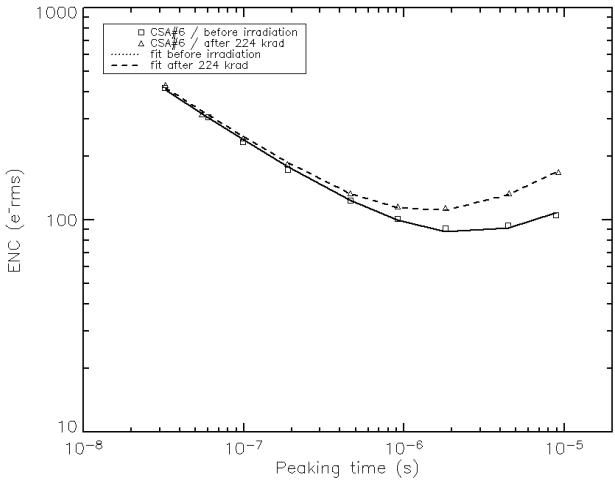

Figure 11: ENC vs. peaking time for the NMOS type CSA#6 before and after 224 krad of irradiation. CSA#6 is polarized with 200 µA current through the input transistor. Lines represent the noise model that fits the data.

Model fitting of the ENC vs. peaking time characteristics for CSA#6 led to the total current responsible for the parallel noise. This total current includes the pad contribution, the compensation current level $il$ and the reset transistor noise. We found $I_{tot}$ ~ 80 pA before irradiation and $I_{tot}$ ~ 160 pA after 224 krad. Finally, by measuring the output signal fall-time on the CSA#6, we found the current through the reset transistor $I_R$ to be 35 pA before irradiation and 2.5 pA after irradiation.

From all that estimations, we derive $I_1 + I_2$ ~ 5 pA for the pad current before irradiation and $I_1 + I_2$ ~ 80 pA for the pad current after irradiation.

The ENC vs. peaking time characteristics shows a 35% augmentation of the $1/f$ noise.

We conclude for the NMOS type CSA#6:

- The main noise source is probably the anti-ESD pad leakage current $I_1+I_2$.
- This pad current generates a negative $I_P$ which obliges us to use a strong compensation current $il$, leading to an even more important parallel noise.
- The pad current is strongly sensitive to the radiations.

## V. CONCLUSIONS

IDeF-X is the very first version of our analogue front-end electronics mainly devoted to Cd(Zn)Te spectro-imaging systems in space. This implies to consider low noise, low-power consumption and radiation tolerance in the design.

We have designed and tested a set of ten very low noise charge sensitive preamplifiers. They have been realized in the standard AMS 0.35 µm CMOS technology. Depending on the type and size of the input transistors, we could obtain a floor noise ranging from 31 to 49 electrons rms, with circuits embedded, but deprived of bonding on the inputs to remove the parasitic noise sources of the test setup. These results have been obtained with 3.3 mW in the CSA stage. Nevertheless, making use of very low dark current detectors, we can obtain almost the same floor noise with long peaking time for shaping, for power consumption down to 165 µW.

We have identified the PMOS type input CSA as the best candidate for future use in a fully integrated spectroscopic chain. As a matter of fact, its noise level is lower than the NMOS type design and works properly without making use of any compensation current source, limiting intrinsically its parallel noise. Consequently, it matches with the low current applications with CdTe detectors at room or moderately low temperatures.

Thanks to these promising results, we have used this chip for hard X-Ray spectroscopy at room temperature with CdTe detectors, leading to a 1.6 keV FWHM spectral resolution at 59.5 keV for power consumption of only 660 µW in the CSA. Compared with the ISGRI detectors and ASIC, it represents a spectral resolution improvement of more than a factor of 3.

Finally, we have irradiated the circuit with a $^{60}$Co source up to 224 krad at 1 krad/h dose rate. We have demonstrated the good tolerance of the design submitted to the total ionizing dose test. We observed a slow increase of the noise level (~30%) after a high dose, much higher than the typical dose in standard space conditions for high energy physics experiments (~ 1 krad/year). We believe that this increase is not mainly due to the CSA design itself, but must be attributed to the standard AMS pads behavior we used in the chip.

The AMS 0.35 µm CMOS technology appeared well adapted to a low noise and low power consumption analogue front end design, devoted to high energy spectroscopy with CdTe. The technology is also tolerant to gamma-rays irradiation. Nevertheless, future designs will imply to design specific pads with lightened anti-ESD protections to limit their contribution to the parallel noise.

## VI. ACKNOWLEDGMENT

The authors wish to thank B. Rattoni from CEA/DRT/FAR/LIST/DETECS for his help during the irradiation tests.

## VII. REFERENCES

[1] F. Lebrun et al., "ISGRI: The INTEGRAL Soft Gamma-Ray Imager", A&A, 411, pp. L141-L148, 2003.
[2] P. Ferrando et al., "SIMBOL-X: a new generation hard x-ray telescope", Proc. SPIE conf. 5168, pp. 65-76, San Diego, Aug. 2003.
[3] P. Von Ballmoos et al., "MAX: a gamma-ray lens for nuclear astrophysics", Proc. SPIE conf., 5168, pp. 482-491, San Diego, Aug. 2003.
[4] S. Schanne et al., "The space borne multi-wave-length gamma-ray burst detector ECLAIRs", Proc. IEEE NSS conf., Rome, 2004.
[5] B. Ravazi, "*Design of analog CMOS integrated circuits*", McGraw-Hill Higher Education, 2001.
[6] V. Radeka, P. O'Connor, "IC Front Ends for Nuclear Pulse Processing", IEEE NSS98 Short Course, Toronto.
[7] C. de La Taille, "Automated system for equivalent noise charge measurements from 10 ns to 10 μs" by Orsay, Nucl. Phys. Proc. Suppl. 32:449-459, 1993.
[8] O. Limousin et al., "The basic component of the ISGRI CdTe gamma-ray camera for space telescope IBIS on board the *INTEGRAL* satellite", Nucl. Inst. and Meth. A, Vol. 428 (1999), pp 216 – 222.
[9] F. Lebrun et al., "INTEGRAL: In flight behavior of ISGRI and SPI", to be published.